# Device Engineering of Perovskite Solar Cells to Achieve Near Ideal Efficiency

Sumanshu Agarwal[1] and Pradeep R. Nair[2]

[1]Department of Energy Science and Engineering, [2]Department of Electrical Engineering

Indian Institute of Technology Bombay, Mumbai, Maharashtra, 400076, India

sumanshu@iitb.ac.in, prnair@ee.iitb.ac.in

*Abstract* — Despite the exciting recent research on perovskite based solar cells, the design space for further optimization and the practical limits of efficiency are not well known in the community. In this manuscript, we address these aspects through theoretical calculations and detailed numerical simulations. Here, we first provide the detailed balance limit efficiency in the presence of radiative and Auger recombination. Then, using coupled optical and carrier transport simulations, we identify the physical mechanisms that contribute towards bias dependent carrier collection, and hence low fill factors of current perovskite based solar cells. Curiously, we find that while Auger recombination is not a dominant factor at the detailed balance limit, it plays a significant role in device level implementations. Surprisingly, our novel device designs indicate that it is indeed possible to achieve efficiencies and fill factors greater than 25% and 85%, respectively, with near ideal super-position characteristics even in the presence of Auger recombination.

**Keywords** —detailed balance limit, fill factor, Auger recombination

# I. INTRODUCTION

The recent reports on organic-inorganic perovskite based solar cells are indeed encouraging with reported efficiencies of the order of 20%[1,2] - a remarkable feat achieved within 3-4 years of active research. Large diffusion length of carriers[3–5] with large extinction coefficient[6] makes such perovskites an ideal material for solar cell application. While a comprehensive literature survey is beyond the scope of this manuscript, we would like to mention that there has been reports of high efficiency solar cells with combination of perovskite materials[2], better reproducibility[7] and stability[8], usage of flexible substrate[9], organic materials as both electron and hole transport layers[10], large grain size perovskite fabrication[11], perovskite based tandem cell[12,13], etc.. There has also been efforts towards better understanding of material properties[14–16], including the recombination strengths[4,17–19] and mobility.[4,19,20]

Despite the above mentioned exciting achievements in experimental research on perovskite solar cells, the corresponding theoretical understanding is lacking in many aspects. For example, while the detailed balance limits of efficiency based on radiative recombination was already reported[21], the effect of Auger recombination is not clearly elucidated. Similarly, the performance limiting factors of current solar cells are yet to be identified, and hence the path for further optimization is ill defined. In this manuscript, we first identify the theoretical performance limits of perovskite solar cells in the presence of radiative and Auger recombination (Section II). Then, through detailed numerical simulations, we identify the physical mechanisms that contribute to sub-optimal performance of current perovskite solar cells (see Fig. 1, also Section III). Finally, we show that near ideal performance can be achieved through appropriate device designs, even in the Auger recombination limit (Section IV).

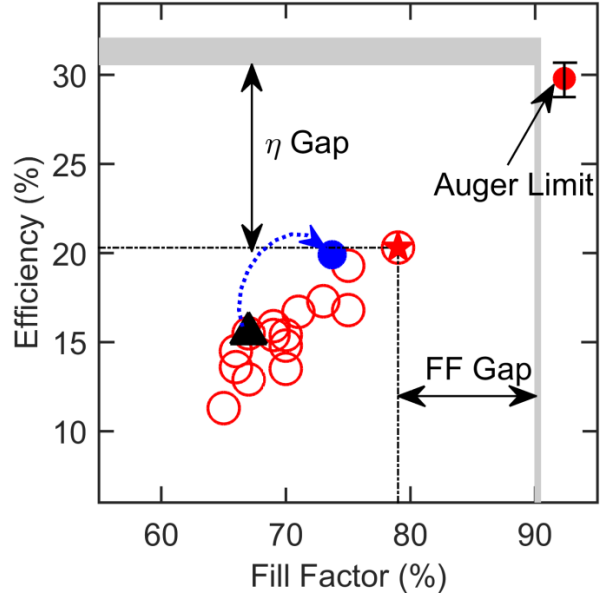

**Fig. 1**: Performance trends of recently reported perovskite solar cells[1,2,26,33] (shown by open circles). The shaded region shows the detailed balance limits (with only radiative recombination) for $FF$ and efficiency for a material with band gap in the range of 1.5-1.6 eV, while the red dot with error bar shows the detailed balance limit in the presence of Auger recombination (assumed band gap of 1.55eV). Lower limit in error bar is corresponding to Auger coefficient $1\times10^{-28}$ cm$^6$s$^{-1}$ while upper limit correspond to Auger coefficient $1\times10^{-30}$ cm$^6$s$^{-1}$. The dotted curve and the solid circle indicate the possible performance improvement in case of valid super-position between dark and light JV characteristics of the data reported in ref.[26]

II AUGER RECOMBINATION AND DETAILED BALANCE LIMITS

The theoretical performance limits of any single junction solar cell are well known in the community since the landmark article by Shockley and Quiesser in 1961[22] and later updated by Green[23] and Tiedje *et al.*[24] These performance limits are dictated by the fundamental recombination mechanisms in the solar cell, as listed below

$$R = \frac{np - n_i^2}{\tau_n(p + p_1) + \tau_p(n + n_1)} + C(np - n_i^2) + (A_n n + A_p p)(np - n_i^2) \quad (1)$$

In the above equation, $R$ denotes the recombination rate, the first term on the RHS denotes trap assisted SRH recombination, the second term denotes the radiative recombination, and the third term denotes the Auger recombination. Detailed balance limit calculations consider the radiative and Auger recombination processes only. For perovskite solar cells, the detailed balance limit for efficiency in the radiative limit is available in literature.[21] However, the effect of Auger recombination is not yet reported. Using the solar spectrum from NREL[25], our calculations indicate that for a material of band gap 1.55eV, the theoretical limit for the efficiency in the presence of only radiative recombination (SQ limit) is 31.45% (with $J_{SC} = 27.3\,\text{mA/cm}^2$, $V_{OC} = 1.28\,\text{V}$, and the $FF = 90.3\%$). While the SQ limits are the theoretical upper bounds for efficiency, Auger recombination could influence the practical efficiency limits and are often computed using the simplifying assumption of $n = p$.[23,24] Under such assumptions, the corresponding performance parameters in the presence of Auger recombination are $J_{SC} = 27.3\,\text{mA/cm}^2$, $V_{OC} = 1.17\,\text{V}$, $FF = 92.3\%$, and $\eta = 29.48\%$ (for a perovskite thickness of 300nm using $A_n = A_p = A = 1 \times 10^{-29}\,\text{cm}^6/\text{s}$, which is very close to reported experimental results[19]). Detailed theoretical calculations of these performance metrics are provided in section I of Supplementary materials. We would like to mention that the above practical efficiency limit in the presence of Auger recombination is very sensitive to the parameter $n_i$, the intrinsic carrier concentration and the Auger recombination coefficient $A$. The intrinsic carrier concentration is unknown for perovskites and we have used $n_i = 9 \times 10^6\,\text{cm}^{-3}$ in our calculations (*i.e.*, based on an assumption of $N_C = N_V = 1 \times 10^{20}\,\text{cm}^{-3}$ and $E_g = 1.55\,\text{eV}$, where $N_C$ and $N_V$ are the effective density of states in the conduction and valence band, respectively). An order of magnitude change in $n_i$, for $A = 1 \times 10^{-29} \text{cm}^6\text{s}^{-1}$, results in approximately 2% change in efficiency (see section II of supplementary material for detail). Similarly, an order of change in $A$, for $n_i = 9 \times 10^6 \text{cm}^{-3}$, leads to ~1% change in efficiency (see figure 1). Hence, accurate experimental estimates of $n_i$ and $A$ is of paramount interest for the community.

### III BIAS DEPENDENT CARRIER COLLECTION AND LOW FF

The achievable limits of efficiency for a solar cell could be much different from the theoretical limits and is dictated by two aspects – (a) the dark JV characteristics $J_{dark}$ and (b) the bias dependent behavior of photo-generated carriers $J_{photo}$. These effects are succinctly captured by the light JV characteristics, $J_{light}$, given by

$$J_{light}(V) = J_{photo}(V) + J_{dark}(V) \qquad (2)$$

The above equation assumes that the dark current remains same during illuminated conditions as well, which need not be universally valid. It is evident from eq. (2) that the optimal performance for a given solar cell can be achieved only when the photo-current is bias independent. For such cases, eq. (2) reduces to the well-known illuminated JV characteristics of a diode where the principle of super-position is valid between the dark and light JV characteristics.

Figure 1 summarizes the efficiency vs. $FF$ trade-off for some of the recently reported high efficiency perovskite solar cells. We also illustrate the detailed balance limit efficiencies and the theoretical FF limits in the same figure. Figure 1 indicates that while the current state of the art perovskite solar cells lag significantly behind the theoretical limits of efficiency, the $FF$ gap is especially intriguing. Further, the physical mechanisms, that contribute toward this loss in efficiency and $FF$, are not well known. More importantly, the practical efficiency limits, as dictated by eq. (2), are also not clearly elucidated.

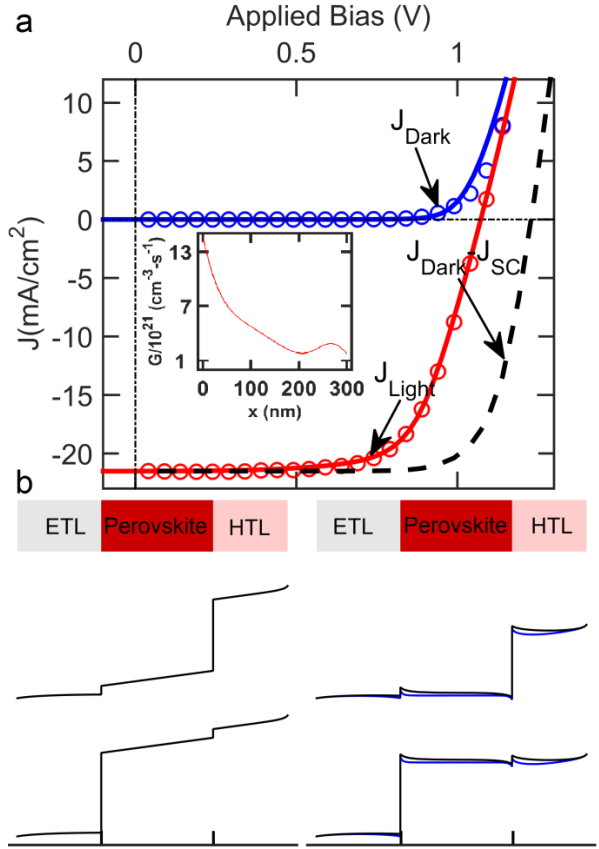

**Fig. 2**: Dark and light JV of the perovskite solar cell as reported by Liu et al.[26] (open symbols). Solid lines are the corresponding simulation results. Dashed curve represents the dark JV downshifted by $J_{SC}$, i.e. expected light JV in case of valid superposition. Simulated photon absorption profile inside perovskite material is shown in inset of part a. Part b represents the energy band diagram of device under equilibrium and at 0.9V in dark (black) and light (blue).

To further understand the practical limits of solar cell efficiency, we performed data analysis using the experimental results reported by Liu et al.[26] Figure 2a shows experimental dark and light JV characteristics (open symbols). Also plotted is the expected JV characteristics based on super-position principle (dashed curve). It is evident from Fig. 2a that the principle of superposition does not hold for these reported devices and the photocurrent is indeed bias dependent. Curiously, we find that if the photo-current were bias independent, the principle of super-position indicates that the upper limit of achievable efficiency for the same device is 20% with a *FF* of 75%. This indicates that bias dependent carrier collection alone results in a loss in efficiency of around 5% for these reported devices. Further, the *FF* from super-position is still much lower than the detailed balance limits (see Fig. 1). This indicates that there is still enough scope to improve the efficiency by better device design that could result in lower dark current (see eq. (2)). Hence it is evident that the current efficiencies could be significantly improved if the physical mechanisms that contribute towards dark current, bias dependent carrier collection and recombination could be identified.

We performed detailed numerical simulations to explore the physical mechanisms that lead to bias dependent carrier collection. Fig. 2a shows a comparison between experimental results (open symbols) from literature[26] and our simulation results (solid lines). The essential physical mechanisms to consider in any solar cell are (a) the carrier generation rate due to optical absorption and (b) the bias dependent carrier collection at appropriate electrodes in the presence of various recombination mechanisms. To address (a), we used the transfer matrix methodology, described by Pettersson *et al.*[27], to estimate the optical absorption and hence the carrier generation rate inside perovskite material. For a thickness of 300 nm for the perovskite layer, our simulations indicate that (see inset of Fig. 2a) the net optical absorption in perovskite material could correspond to a $J_{SC} \sim 21.5$ mA/cm$^2$ (under the assumption that all photo-generated carriers are collected at short circuit conditions[28]), which is very close to the experimental results reported by Liu *et al.*[26] With calibrated estimates for carrier generation rates, we addressed the carrier collection through self-consistent solution of continuity and Poisson's equations.[29,30] We neglect the role of excitons, whose effects were shown to be negligible at low illumination intensities[17] (*i.e.* 1 Sun). Our simulation methodology and parameters are calibrated with experimental results, as reported previously[31], and the details are provided in Section IV and V of supplementary materials.

In self-consistent simulation of Poisson and carrier continuity equations, we explicitly considered all the recombination mechanisms listed in eq. (1). Specifically, we assumed temperature independent auger recombination (see eq. 1), with $A_n = A_p = 1 \times 10^{-29}$ cm$^6$s$^{-1}$, a value very close to that literature reports.[19] The radiative recombination rate estimated theoretically using Van Roosbroeck model[32] ($C = 2 \times 10^{-13}$ cm$^3$s$^{-1}$, see section III of supplementary material) vary significantly from the corresponding rates obtained from Photoluminiscence decay[17,18] ($C \sim 1 \times 10^{-10}$ cm$^3$s$^{-1}$). Hence, we assumed $C = 3 \times 10^{-11}$ cm$^3$s$^{-1}$ in our numerical simulations. The experimentally observed dark current ideality factor is close to 2, which indicates significant trap assisted recombination. This allows one to estimate the minority carrier lifetime (assuming $\tau_n = \tau_p = \tau$ see eq. (1)) from the relation $V_{oc} = 2kT/q \, ln(J_{sc}/J_0)$, where $J_0 \sim n_i W / 2\tau$, W being the thickness of perovskite layer (see ref.[29]). Using the experimental results for $V_{OC} = 1.07$ V, $J_{SC} = 21.5$ mAcm$^{-2}$, and $W = 300$ nm, the above relations indicate that $\tau$ is the order of $10^{-6}$ s for perovskite. Note that this estimate is very close to that obtained from PL measurements.[4] Through detailed simulations, we find that $\tau = 2.73 \times 10^{-6}$ s and $\mu = 0.2$ cm$^2$/Vs (where $\mu$ is the mobility of charge carrier in the perovskite) along with the before mentioned parameters for radiative and Auger recombination can anticipate all the relevant features of experimental dark and light IV. The assumed value for mobility along with the above mentioned minority carrier lifetime gives a diffusion length of around 1µm, close to the reported experimental results.[3]

Interestingly, our simulations could reproduce all relevant features of $J_{dark}$, which includes a diode ideality factor of 2 indicating trap assisted process as the dominant recombination mechanism and a large bias voltage exponent of 2 which indicates the space charge limited transport (corresponding parameters are extracted in section V of supplementary materials and ref.[31]). In addition to this, a perfect agreement between numerical simulations and experimental results is observed for the light JV characteristics for all relevant parameters like $J_{SC}$, $FF$, and $V_{OC}$, thus validating the simulation methodology and the material parameters used. The energy band diagrams shown in Fig. 2b provide important insights towards the bias dependence of photocurrent. It is evident that the high band offsets at ETL/perovskite junction and perovskite/HTL junction act as near ideal blocking contacts. Equilibrium band diagram shown in fig. 2b suggest that till $V = V_{bi}$ ($V_{bi}$ is the built in potential), the photogenerated carriers will be collected effectively by the contact layers. When applied bias is more than $V_{bi}$, the electric field in the perovskite and ETL/HTL layers is not favorable for carrier collection at desired contacts. This bias dependent carrier collection in turn leads increased recombination of photo-generated carriers in perovskite layer. As a result the photocurrent shows a bias dependence thus limiting the efficiencies with low $FF$.

## IV PRACTICAL EFFICIENCY LIMITS

The results shown in Fig. 2 indicate that the bias dependence of photocurrent is due to the fact that the collection of photo-generated carriers is mainly transport limited. It is now evident that to reduce the bias dependence of photocurrent, one should ensure that carrier transport through the entire device is also bias independent – *i.e.*, the electric field assisted drift component should be reduced while increasing the diffusive carrier transport. The E-B diagram in Fig. 2 indicates that significant bias dependent carrier transport occurs in both the perovskite as well as the contact layers. Increasing the mobility of carriers in the contact layers is expected to improve the $FF$, however, this might not be enough to reduce the bias dependence (see Section VI of Supplementary materials for a detailed discussion on this). Curiously, the carrier collection can be made effectively bias independent if the various layers are doped appropriately. We will now explore the performance improvement due to these schemes.

Figure 3a provides a few schemes to reduce the bias dependence of photo-current. Scheme S1 involves doping of perovskite to negate the effects of bias dependent transport of photo-generated carriers. Scheme S2 attempts to reduce the bias dependent carrier collection by doping the ETL/HTL, while scheme S3 explores the option of staggered doping profiles in perovskite along with doping in ETL/HTL. Figure 3b indicates the efficiency vs. $FF$ landscape for the scheme S2. The performance trends for schemes S1 and S3 are provided in the supplementary materials (Fig SF6). The solid symbol A in Fig. 3b denotes the efficiency of the device structure shown in Fig. 2 with $\tau = 2.73 \times 10^{-6}$ s, radiative recombination coefficient $C = 3 \times 10^{-11}$ cm$^3$s$^{-1}$, and Auger recombination coefficient $A_n = A_p = 1 \times 10^{-29}$ cm$^6$s$^{-1}$. Curve A to B is the performance improvement through scheme S2, *i.e.*, contact layer doping. Scheme S2 provides considerable improvement in efficiency and $FF$ with increase in ETL/HTL doping (ranging from $1 \times 10^{15}$ cm$^{-3}$ to $1 \times 10^{19}$ cm$^{-3}$). Note that this scheme yields better performance through improvements in $FF$, as the contact layer doping could reduce the bias dependent carrier collection. An apparent saturation in performance improvement was observed using scheme S2 at ~84% $FF$ when ETL/HTL doping density is of the order of $10^{18}$ cm$^{-3}$, with valid super-position between dark and light IV. The doping levels and the corresponding performance parameters are provided as Table S2 and S3 in supplementary materials.

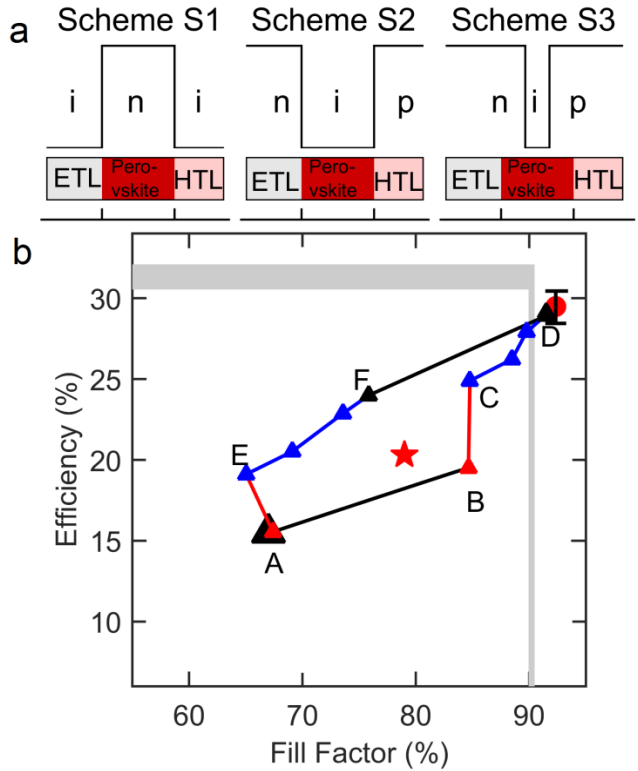

Fig. 3: Device engineering to improve the efficiency of perovskite solar cells. Part (a) shows different schemes (S1, S2, and S3) to reduce the bias dependence of photocurrent while part (b) shows the $FF$ and Efficiency trends for Scheme S2. In part (b) the solid circle denotes the detailed balance limit while the solid star denotes the highest efficiency reported in literature.[2] Point A denotes the base case efficiency (i.e., the results shown in Fig. 2), while B denotes the efficiency of the same device with heavy contact layer doping. Curve AE and BC denotes the corresponding performance improvement due to ideal optics, while curve EF and CD denotes further improvement due to better quality perovskite material. Note that the best achievable efficiency is about 0.4% lower than the detailed balance limit in the presence of auger recombination.

Figure 3b also shows some interesting trends on further device optimization. For example, curve A-E indicates the improvement that can be obtained to the base case device with ideal optics (*i.e.*, through front surface texturing, anti-reflection coating, appropriate materials for the ETL to reduce undesired absorption/reflection, etc., which could yield in $J_{SC} = 27.3 \, \text{mA/cm}^2$) while the curve E-F indicates the further performance improvement due to reduced SRH recombination (with $\tau_{SRH} = 2.73 \, \text{ms}$ a value comparable to that of solar grade silicon), and with $C = 3 \times 10^{-13}$ cm$^3$/s – the theoretical estimate for radiative recombination rate. Now, curve F-D shows the further improvement in efficiency with scheme S2, *i.e.*, contact layer doping. Interestingly, we find that the best achievable efficiencies are about ~0.4% lower than the theoretical limits ($J_{SC} = 27.3 \, \text{mAcm}^{-2}$, $V_{OC} = 1.16 \, \text{V}$, $FF = 91.5\%$, $\eta = 29.1\%$). This difference is due to the fact that the fundamental assumption of $n = p$ used in Auger limit calculations is not valid under maximum power point conditions (see Section VII of supplementary materials), thus leading to increased recombination and hence lower efficiency in device level implementations.

Now, we would like to explore the implications of perovskite doping on the above mentioned performance trends for the device with best efficiency (point D in Fig. 3b). An increase in perovskite doping can affect the performance in two ways – (a) an undesired band bending at ETL/perovskite or perovskite/HTL interface, and (b) increase in Auger recombination (see eq. (1)). For efficient charge transfer from perovskite to transport layer, the Fermi level in perovskite should not be above (below) Fermi level of ETL (HTL) in equilibrium. This indicates that if $n_P$ is the doping in perovskite and $n_E$ is doping in ETL then $\frac{n_E}{n_P} \geq \frac{N_{CE}}{N_{CP}} \exp(\frac{\Delta E_C}{kT})$, where $N_{CP} = N_{CE} = 1 \times 10^{20}$ cm$^{-3}$ are the effective density of states for electrons in perovskite and ETL, respectively and $\Delta E_C$ is the conduction band offset between perovskite and ETL. For $\Delta E_C = 0.2$ eV and ETL doping of $1\times10^{19}$ cm$^{-3}$, the above estimate indicates that a performance drop (mainly through $J_{SC}$) is expected for perovskite doping greater than $1\times10^{16}$ cm$^{-3}$, which is observed in detailed numerical simulations. Similar arguments hold good for perovskite/HTL interface if the perovskite is p-type doped. Moreover, increase in perovskite doping leads to an increase Auger recombination as well. Eq. (1) indicates that Auger recombination dominates the radiative recombination if the carrier density is such that $An > C$. Using $C = 3\times10^{-13}$ cm$^3$s$^{-1}$ and $A = 1\times10^{-29}$ cm$^6$s$^{-1}$, we again find that perovskite doping on the higher side of $1\times10^{16}$ cm$^{-3}$ could result in reduced efficiency. As expected, these trends are also supported by detailed numerical simulations.

Finally, we would like to mention that the practical limits of efficiency discussed in this manuscript depend on several parameters. The most dominant among them is $n_i$, the intrinsic carrier concentration whose accurate estimates are yet to be reported in literature. Hence it is very essential that various parameters such as recombination coefficients and intrinsic carrier concentration be explored through multiple experimental techniques such that the practical limits are further well defined.

## VI Conclusion

In this manuscript, we provide a comprehensive modeling framework to understand and optimize the performance of Perovskite based solar cells. Our theoretical analysis and numerical simulations identify (a) the detailed balance performance limits (b) the physical mechanisms that contribute to sub-optimal performance of current perovskite based solar cells, and (c) suggest novel schemes to further improve the performance. Indeed, our simulations show that it is possible to achieve > 25% efficiency with near ideal FF for an optimally designed perovskite based solar cell – a result that could be of immense interest to the community.

**Acknowledgements:** This paper is based upon work supported in part by the Solar Energy Research Institute for India and the United States (SERIIUS), funded jointly by the U.S. Department of Energy (under Subcontract DE-AC36-08GO28308) and the Govt. of India's Department of Science and Technology (under Subcontract IUSSTF/JCERDC-SERIIUS/2012). The authors also acknowledges Center of Excellence in Nanoelectronics (CEN) and National Center for Photovoltaic Research and Education (NCPRE), IIT Bombay for computational facilities.

**Author Information:** All correspondence should be addressed to S.A. at sumanshu@iitb.ac.in or P.R.N. at prnair@ee.iitb.ac.in. Authors declare no competing financial interest.

Supplementary material for

# Device Engineering of Perovskite Solar Cells to Achieve Near Ideal Efficiency

Sumanshu Agarwal[1] and Pradeep R. Nair[2]

[1]Department of Energy Science and Engineering, [2]Department of Electrical Engineering
Indian Institute of Technology Bombay, Mumbai, Maharashtra, 400076, India
sumanshu@iitb.ac.in, prnair@ee.iitb.ac.in

Correspondence should be addressed to:

[*]Sumanshu Agarwal, Department of Energy Science and Engineering, IIT Bombay, Powai, Mumbai-400076, email: sumanshu@iitb.ac.in

[*]Prof. Pradeep R. Nair, Department of Electrical Engineering, IIT Bombay, Powai, Mumbai-400076, email: prnair@ee.iitb.ac.in

**Section I: Detailed balance limit for single active layer solar cell**

Detailed balance limit for solar cell was first proposed by Shockley et al.[1] and practical limits were further discussed by Green[2] and Tiedje et al.[3] Using the solar spectrum[4] shown in figure SF1 we calculated the limiting values for different parameters of solar cell as a function of bandgap. For the calculation of limits we assumed that radiative process is the only recombination mechanism in the device as it is fundamental and cannot be avoided if body is at nonzero temperature. Corresponding plots for the limits are shown in fig SF2.

This detailed balance limit estimate is based on the assumption that solar cell is a perfectly black body for photons of energy more than band gap and white body for photons with energy less than band gap. Based on this assumption we calculated the total number of photons absorbed in the cell which in turn is equal to the generation rate of electron-hole pairs. If $S(E)$ is the photon flux available in solar spectrum for energy E in the energy interval $dE$ then total number of photons above band gap energy, $Ph(E_g)$, will be,

$$Ph(E_g) = \int_{E_g}^{E_\infty} S(E)dE . \tag{S1}$$

All these photons are absorbed by the cell and generate an equal number of electron-hole pairs.

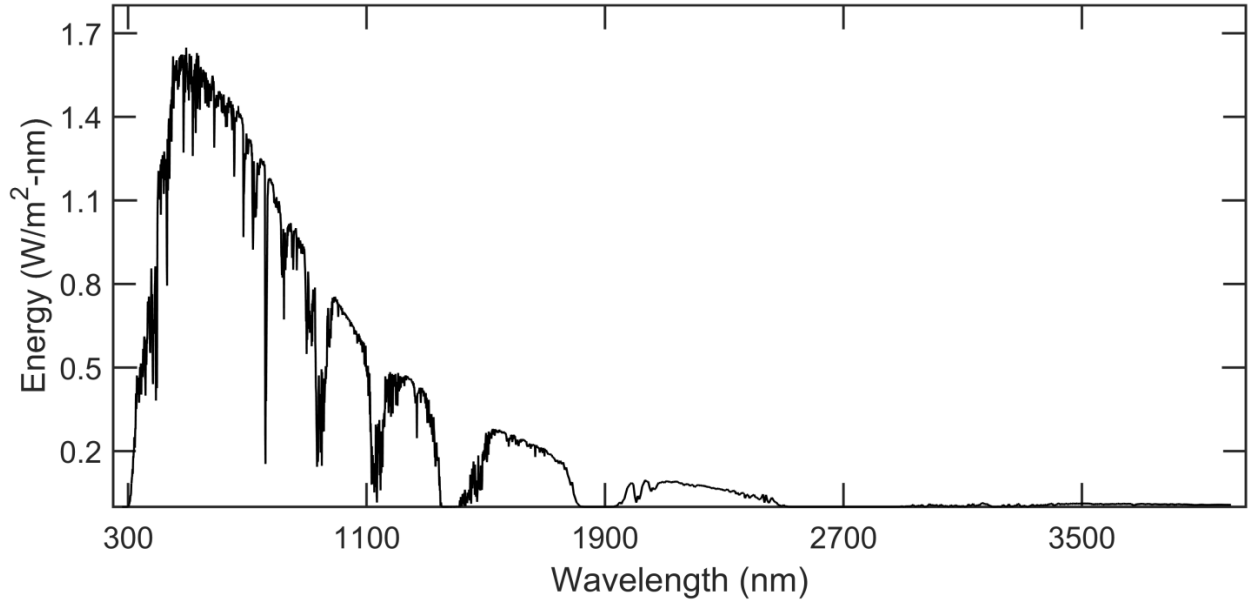

**Fig SF1**: Global solar irradiance as reported by NREL.

Radiative recombination under black body radiation limit depends on semiconductor temperature and separation of quasi Fermi levels.[5] Any photogenerated e-h pair which does not contribute to the recombination will be extracted from the cell in the form of current. So net current density, $J$, obtained from the cell will be

$$J(E_g,V) = q(Ph(E_g) - RR(E_g,V)) \tag{S2}$$

In the above equation $RR$ is the radiative recombination flux and $V$ is the applied bias across the cell. Radiative recombination has two components: one is band gap dependent and other is bias dependent. The black body radiation from the cell at thermal equilibrium, which is equivalent to radiative recombination at thermal equilibrium, is given by:

$$RR_0(E_g) = \frac{2\pi}{c^2h^3} \int_{E_g}^{\infty} \frac{E^2 dE}{\exp(E/kT)-1} \text{ cm}^{-2}\text{s}^{-1} \tag{S3}$$

Radiative recombination at applied bias V depends on the split in quasi Fermi level[5] and given by equation S4.

$$RR(E_g,V) = \frac{2\pi}{c^2h^3} \int_{E_g}^{\infty} \frac{E^2 dE}{\exp((E-qV)/kT)-1} \text{ cm}^{-2}\text{s}^{-1} \tag{S4}$$

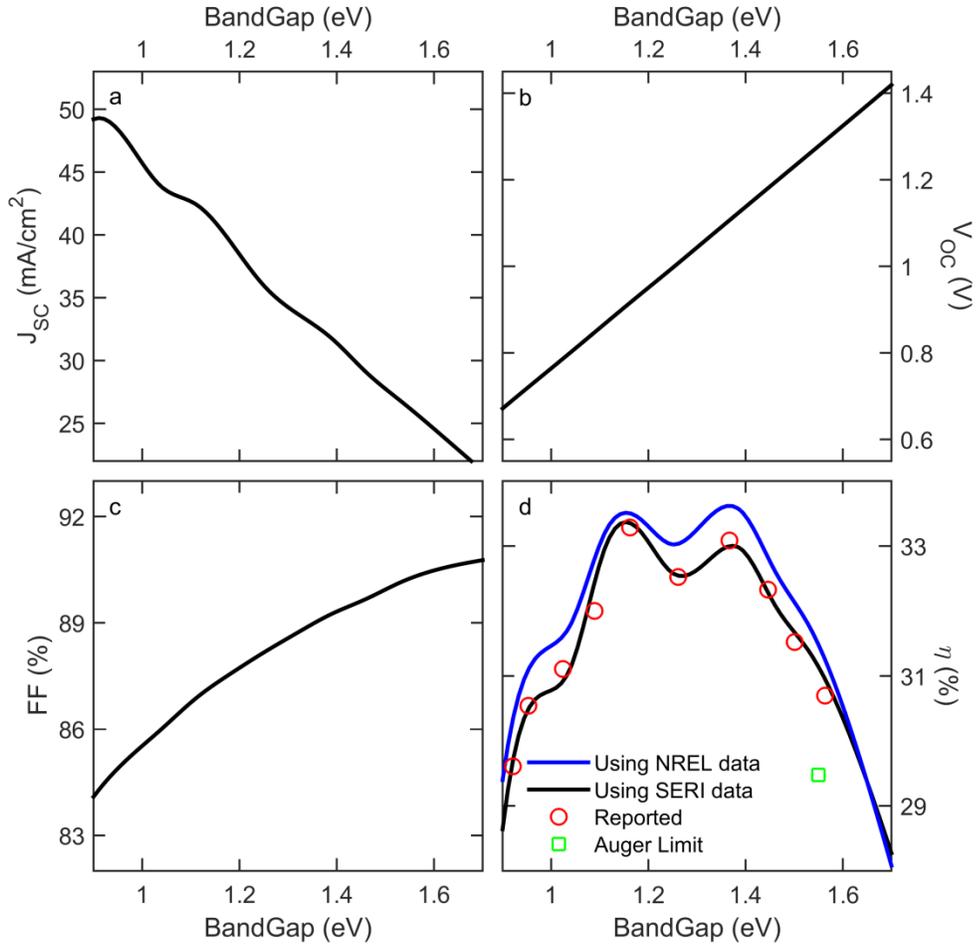

**Fig SF2**: Detailed balance limit for (a) $J_{SC}$, (b) $V_{OC}$, (c) FF, and (d) Efficiency of single bandgap active layer solar cell calculated using NREL data.

For any semiconductor if applied bias is less than $V_{OC}$ and band gap is more than ~ 0.2 eV, then $qV \ll E$ on the scale of $kT$. Therefore eq. S4 can be reduced to

$$RR(E_g, V) = RR_0(E_g)\exp(qV/kT) \tag{S5}$$

When cell is in thermal equilibrium then recombination is balanced by generation in the cell. Therefore net recombination in the cell more accurately can be written as

$$RR(E_g, V) = RR_0(E_g)\exp(qV/kT) - RR_0(E_g) \tag{S6}$$

Using eq. S2 and S6, illuminated JV for the cell can be given by

$$J(E_g, V) = q(Ph(E_g) - RR_0(E_g)(\exp(qV/kT) - 1)) \tag{S7}$$

Maximum available short circuit current density, $J_{SC}(E_g)$, for a semiconductor with band gap $E_g$ is $qPh(E_g)$ while open circuit voltage $V_{OC}(E_g)$ is the voltage when $J(E_g, V) = 0$. Therefore maximum available $V_{OC}(E_g)$ can be given by,

$$V_{OC}(E_g) \sim \frac{kT}{q}\ln\left(\frac{Ph(E_g)}{RR_0(E_g)} + 1\right), \tag{S8}$$

which is purely a function of bandgap and temperature for a given solar flux. Power ($P$) generated by cell is the product of $J$ and $V$. Maximum value of power ($P_{MPP}$) was calculated numerically. Ratio of $P_{mpp}$ to the product of $J_{SC}(E_g)$ and $V_{OC}(E_g)$ is the fill factor ($FF(E_g)$) of the cell. Efficiency of the cell is given by ratio of $P_{mpp}(E_g)$ to the power incident on the cell ($P_{in}$). The detailed balance limit estimates are shown in Fig SF2. $J_{SC}$, $V_{OC}$, and $FF$ as a function of band gap of semiconductor under radiative limit are shown in part (a), (b), and (c) respectively. Part (d) shows the variation of efficiency as a function of band gap. Here we provide two estimates, one using the latest NREL[4] data for solar spectrum and the other using SERI[6] data, which was used by Tiedje *et al.*[3] The estimates from Tiedje *et al* are also provided in the same figure (symbols). This shows that our calculations reproduce literature results for the same input solar spectra. For the main manuscript, we have used calculations based on the NREL data.

**Section II: Effect of Auger recombination**

We also performed auger recombination limit study using the method provided in Tiedje *et al.*[3] For 300 nm thick active layer perovskite solar cell Auger recombination brings the efficiency down by ~2% (see fig SF2 part d). A similar analysis for Si solar cell was done by Green[2] and for GaAs solar cells was reported by Sandhu *et al*[7]. The Auger recombination was calculated under the assumption that $n = p$. The auger recombination rate in the device will be given by

$$RA = \int_0^W A(np - n_i^2)(n + p)dx \tag{S9}$$

Where $W$ is the thickness of active layer, $A$ is the auger recombination coefficient (assuming electron and hole Auger recombination coefficients are same), $n$ and $p$ are the electron and hole concentration respectively. Under Boltzmann's approximation, product of $n$ and $p$ as a function of voltage is given by

$$np = n_i^2 \exp(\frac{qV}{kT}) \tag{S10}$$

When $n = p$ and $np \gg n_i^2$ eq. S9 reduces to

$$RA = 2An_i^3 \exp(\frac{3qV}{2kT})W \tag{S11}$$

Equation S11 is used to get the Auger recombination flux and net $J-V$ characteristics is given by

$$J(E_g, V) = q(Ph(E_g) - RR(E_g, V) - RA(E_g, V)) \tag{S12}$$

The new set of solar cell parameters can now be calculated using the $J-V$ characteristics obtained from eq. S12.

For $n_i = 9 \times 10^6 \, \text{cm}^{-3}$ and $A = 1 \times 10^{-29} \, \text{cm}^6 \text{s}^{-1}$, the performance parameters are $J_{SC} = 27.3 \, \text{mA/cm}^2$, $V_{OC} = 1.17 \, \text{V}$, $FF = 92.34\%$, $\eta = 29.48\%$. One order increase in $n_i$ reduces the efficiency to 26.31% with $J_{SC} = 27.3 \, \text{mA/cm}^2$, $V_{OC} = 1.05 \, \text{V}$, and $FF = 91.73\%$, while one order decrease in $n_i$ increases the efficiency to 31.4% with $J_{SC} = 27.3 \, \text{mA/cm}^2$, $V_{OC} = 1.27 \, \text{V}$, and $FF = 90.75\%$. Similarly for $n_i = 9 \times 10^6 \, \text{cm}^{-3}$, one order change in $A$ from $1 \times 10^{-29} \, \text{cm}^6 \text{s}^{-1}$ results 1% change in efficiency mainly through change in $V_{OC}$.

**Section III: Theoretical estimate for radiative recombination coefficient**

Detailed numerical simulation of the performance of a solar cell requires the knowledge of recombination parameters for the material used in the cell. Hence for a perovskite solar cell, we need an estimate of the radiative recombination coefficient. We extracted this value theoretically using the well-known Van Roosbroeck[8] model. According to this model, at thermal equilibrium, rate of radiative recombination of electron-hole pair should be equal to rate of generation of electron-hole pairs by thermal radiation. Therefore radiative recombination in thermal equilibrium can be written as

$$R = \int P(v)\rho(v)dv$$

Where $P(\nu)$ is the probability that a photon of frequency $\nu$ in the frequency interval $d\nu$ will be absorbed in the unit volume and $\rho(\nu)$ is the density of photons with frequency $\nu$ available in that unit volume.

Roosbroeck et al.[8] has shown that above equation can be simplified to

$$R = 1.785 \times 10^{22} (T/300)^4 \int_{E_g}^{\infty} \frac{n^2 \kappa u^3 du}{e^u - 1} \, cm^{-3} s^{-1}$$

$T$ = temperature on absolute scale
$u = h\nu / kT$
$n$ = refractive index
$\kappa$ = extinction coefficient
$E_g$ = band gap of material

Using this equation radiative recombination can be calculated at thermal equilibrium. Also $Cn_i^2$ gives the radiative recombination in thermal equilibrium[9] where $C$ is the radiative recombination coefficient and $n_i$ is intrinsic carrier concentration in the semiconductor. Using reported values of refractive index and extinction coefficient for perovskite[10] with $n_i$ of the order of $9 \times 10^6 \, cm^{-3}$ we obtained $C = 2 \times 10^{-13}$ cm$^3$/s.

**Section IV: Simulation methodology and parameters used for simulation**

To explore the dark and light JV characteristics, self consistent numerical solution of continuity, charge transport, and Poisson's equation was done using synopsis device simulation tool, Sentaurus[11]. The generalized form of these equation are given below, where $\xi$ stands for charge particle (electron or hole).

*Continuity equation:* This equation accounts the effect of generation ($G$) and recombination ($R$) of carrier density at point $x$ inside the cell on the current density ($J$). Mathematically, the steady state form of continuity equation can be written as,

$$\pm \frac{1}{q} \frac{dJ_\xi}{dx} = G_\xi(x) - R_\xi(x) \tag{S13}$$

Negative sign in equation S13 stands for electron and positive sign for holes. $q$ is the charge on electron

*Charge transport equation:* This equation connects the diffusive and drift component of current density.

$$\frac{1}{q} J_\xi = \pm D \frac{dn_\xi}{dx} + \mu_\xi n_\xi E \tag{S14}$$

Negative sign in equation S14 stands for holes and positive sign for electrons.

*Poisson's equation:* The effect of charge carrier concentration on the electric field/electrostatic potential is governed by Poisson's equation given below.

$$\frac{dE}{dx} = \frac{q}{\varepsilon \varepsilon_0} \rho(x) \qquad (S15)$$

Where $\rho(x)$ is the density of charge carrier at point $x$. The parameter space used for the solution of these equation is provided in Table S1.

**Table S1**: Parameters used for the simulation of perovskite solar cell

| Parameter | Symbol | Unit | Perovskite | ETL | HTL |
|---|---|---|---|---|---|
| Electron affinity | $\chi$ | eV | -3.8 | -4.0 | -2.15 |
| Band gap | $E_g$ | eV | 1.55 | 3.2 | 3.0 |
| Effective DOS for electron | $N_C$ | $cm^{-3}$ | $1\times 10^{20}$ | $1\times 10^{20}$ | $1\times 10^{20}$ |
| Effective DOS for hole | $N_V$ | $cm^{-3}$ | $1\times 10^{20}$ | $1\times 10^{20}$ | $1\times 10^{20}$ |
| Mobility of electron | $\mu_e$ | $cm^2/V$-s | 0.2 | 0.017 | $0.04^*$ |
| Mobility of hole | $\mu_h$ | $cm^2/V$-s | 0.2 | $0.017^*$ | 0.04 |
| SRH lifetime | $\tau_e = \tau_h$ | s | $2.73\times 10^{-6}$ | - | - |
| Radiative recombination coefficient | $C$ | $cm^3 s^{-1}$ | $3\times 10^{-11}$ | - | - |
| Auger recombination coefficient | $A_n = A_p$ | $cm^6 s^{-1}$ | $1\times 10^{-29}$ | - | - |
| Contact work-function | $\phi$ | eV | - | -4.2 | -4.95 |

[$^*$Note: electron mobility in HTL or hole mobility in ETL are chosen arbitrarily. However because of large barriers carriers cannot crossover and these values do not have any effect on simulation results.]

### Section V: ideality factor and voltage exponent of base case simulated device

We have previously shown that perovskite solar cell exhibit two universal features, a) diode ideality factor is two at low bias and b) voltage exponent is two at high bias[12]. We find that in present case also, where radiative recombination and Auger recombination have also been included, same characteristic features are still available (Fig. SF3).

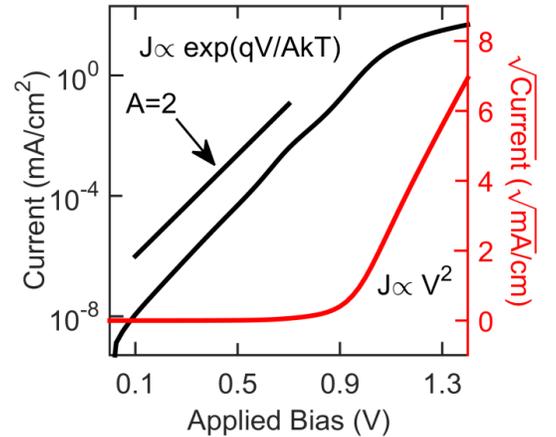

**Fig SF3:** base case device shows low bias ideality factor 2 and space charge limited current at high bias in dark

**Section VI: Effect of contact layer mobility on the performance of device**

As discussed in the main text that performance of the cell can be improved by using high mobility contact layers, our detailed numerical simulation also indicates the same (Fig SF4). It is clear that increasing the mobility of ETL/HTL could improve the performance, but it is still far below the detailed balance limit. Superposition also does not hold if mobility values for carriers are high in contact layers but not doped. Also if the charge carrier mobility in contact layers is more than carrier mobility in perovskite, then the performance does not improve any further. Figure SF4 shows the improvement in device performance when e-h mobility in ETL and HTL are increased.

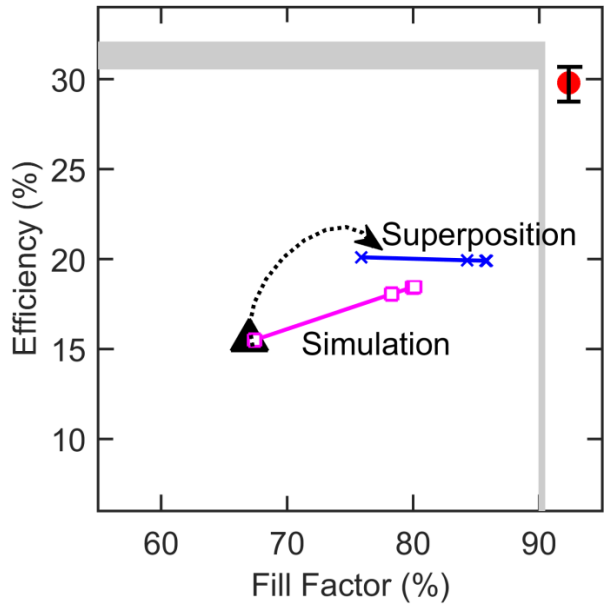

**Fig SF4:** Effect of mobility of contact layers on the performance of the cell. Failure of superposition can be observed by the difference in curve obtained from simulation (line with squares) and superposition (line with cross).

**Section VII Difference in analytical and simulation practical limit of efficiency**

As discussed in the main text, the assumption of $n = p$ is not valid near the maximum power point conditions (see Fig. SF5). It leads to higher recombination in practical device as compared to the detailed balance analysis and reduces the open circuit voltage. Figure SF5b shows a comparison of the generation rate and recombination rate (from both detailed balance analysis and simulation) integrated over the thickness of perovskite layer. It is evident that the recombination in an actual device is more as compared to same in detailed balance analysis. As a consequence, the $V_{OC}$ of the device is a lower than the corresponding detailed balance limit. In addition, voltage at maximum power point ($V_{MPP}$) is also different, which are indicated by points A (1.087 V from numerical simulations) and B (1.097 from detailed balance analysis) in figure

SF5b. Increased recombination in simulation lead to some difference in $J_{MPP}$ also. However, this difference is very small but it also contributes to efficiency difference. The inset of SF5b clearly indicates that net recombination in simulation result is little higher than analytical result. Consequently the $J_{MPP}$ for simulation is lower than analytical $J_{MPP}$. The efficiency of solar cell is given by

$$\eta = \frac{J_{MPP} \times V_{MPP}}{P_{in}} \times 100$$

Differentiating both side lead to

$$\frac{\Delta \eta}{\eta} = \frac{\Delta J_{MPP}}{J_{MPP}} + \frac{\Delta V_{MPP}}{V_{MPP}}$$

$$\Rightarrow \Delta \eta = \eta_{ana} \left( \frac{\Delta J_{MPP}}{J_{MPP_{ana}}} + \frac{\Delta V_{MPP}}{V_{MPP_{ana}}} \right) \quad (S16)$$

The subscript *ana* in equation (S16) corresponds to analytical result and $\Delta$ represents the difference in analytical and simulation result. Analytical efficiency of device is 29.48% with $J_{MPP}$ and $V_{MPP}$ being 26.87 mA/cm$^2$ and 1.097 V respectively. Corresponding simulation results are 26.76 mA/cm$^2$ and 1.087 V respectively. Equation (S16), when solved using these values, yields $\Delta \eta \sim 0.4\%$, which is exactly the difference observed in detailed numerical simulations and analytical calculations.

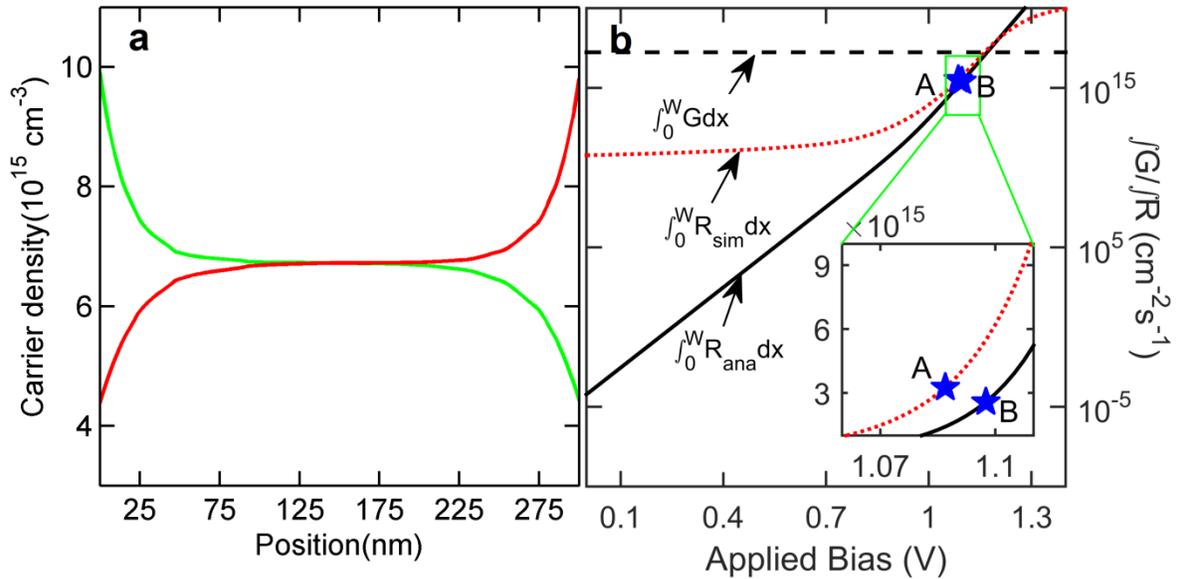

**Figure SF5**: Difference in recombination in analytical approach and detailed simulation. Part a shows the carrier concentration profile inside perovskite for device with optimum efficiency at 1.05 V applied voltage. Recombination profile integrated over perovskite thickness along with integrated carrier generation rate as a function of applied bias is shown in part b.

## Section VIII: Performance metrics for different device schemes discussed in main text

Performance change using scheme S1 and S3 are shown in figure SF6. Curve AX shows the performance change using scheme S1 while change for scheme S3 is shown by curve BY. It can be seen that although schemes S1 and S2 were supposed to have potential to improve the performance but improvement is very less for low doping density while high doping degrades the performance.

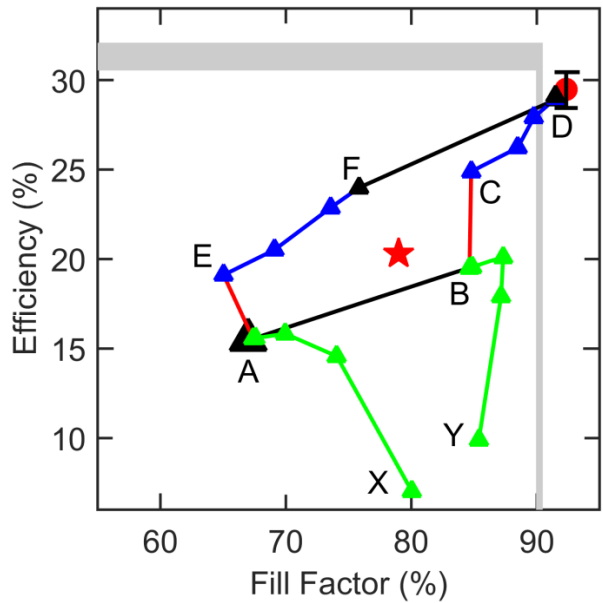

**Figure SF6**: Performance tradeoff using different proposed scheme discussed in the main text. Curve AX shows the performance variation if scheme S1 is used while BY shows the performance trend if scheme S3 is used. Other curves are discussed in main text.

**Table S2**: Parameters for different nodal points in Figure 3b

| Point → | A | B | C | D | E | F |
|---|---|---|---|---|---|---|
| C (cm$^3$s$^{-1}$) | $3\times10^{-11}$ | $3\times10^{-11}$ | $3\times10^{-11}$ | $3\times10^{-13}$ | $3\times10^{-11}$ | $3\times10^{-13}$ |
| $\tau_e = \tau_h$ (s) | $2.73\times10^{-6}$ | $2.73\times10^{-6}$ | $2.73\times10^{-6}$ | $2.73\times10^{-3}$ | $2.73\times10^{-6}$ | $2.73\times10^{-3}$ |
| Doping value of contact layers (cm$^{-3}$) | - | $1\times10^{19}$ | $1\times10^{19}$ | $1\times10^{19}$ | - | - |
| G (cm$^{-3}$s$^{-1}$) | $4.48\times10^{21}$ | $4.48\times10^{21}$ | $5.67\times10^{21}$ | $5.67\times10^{21}$ | $5.67\times10^{21}$ | $5.67\times10^{21}$ |

**Table S3**: Performance metrics for different nodal points in Figure 3b

| Point → | A | B | C | D | E | F |
|---|---|---|---|---|---|---|
| $J_{SC}$ | 21.5 | 21.5 | 27.3 | 27.3 | 27.3 | 27.3 |
| $V_{OC}$ | 1.07 | 1.07 | 1.08 | 1.16 | 1.08 | 1.16 |
| FF | 67.4 | 84.6 | 84.8 | 91.5 | 65.0 | 75.8 |
| $\eta$ | 15.5 | 19.5 | 24.9 | 29.1 | 19.1 | 24.0 |

**References:**

[1] S. William and J.Q. Hans, J. Appl. Phys. **32**, 510 (1961).

[2] M.A. Green, IEEE Trans. Electron Devices **31**, 671 (1984).

[3] T. Tiedje, E. Yablonovitch, G.D. Cody, and B.G. Brooks, IEEE Trans. Electron Devices **31**, 711 (1984).

[4] Solar Spectral Irradiance: ASTM G-173, NREL, < http://rredc.nrel.gov/solar/spectra/am1.5/ASTMG173/ASTMG173.html>.

[5] W. Ruppel and P. Wurfel, IEEE Trans. Electron Devices **27**, 877 (1980).

[6] R. Matson, R. Bird, and K. Emery, *Terrestrial Solar Spectra, Solar Simulation, and Solar Cell Efficiency Measurment* (Golden, Co, 1981).

[7] S. Sandhu, Z. Yu, and S. Fan, Opt. Express **21**, 1209 (2013).

[8] W. van Roosbroeck and W. Shockley, Phys. Rev. Lett. **94**, 1558 (1954).

[9] R.N. Hall, in *Proc. IEE - Part B Electron. Commun. Eng.* (1959), pp. 923–931.

[10] C.-W. Chen, S.-Y. Hsiao, C.-Y. Chen, H.-W. Kang, Z.-Y. Huang, and H.-W. Lin, J. Mater. Chem. A (2014).

[11] Sentaurus Device Simulation Tool, Synopsys, (2011).

[12] S. Agarwal, M. Seetharaman, N.K. Kumawat, A.S. Subbiah, S.K. Sarkar, D. Kabra, M.A.G. Namboothiry, and P.R. Nair, J. Phys. Chem. Lett. **5**, 4115 (2014).